\documentclass{article}
\usepackage{spconf,amsmath,graphicx}
\usepackage{booktabs}
\usepackage{multirow}
\usepackage{graphicx}

\title{LEARNING EMOTION-INVARIANT SPEAKER REPRESENTATIONS FOR SPEAKER VERIFICATION}
\name{Jingguang Tian, Xinhui Hu, Xinkang Xu}
\address{Hithink RoyalFlush AI Research Institute, Zhejiang, China}
\begin{document}
\ninept
\maketitle
\begin{abstract}
\vspace{-0.1cm}
In recent years, the rapid progress in speaker verification (SV) technology has been driven by the extraction of speaker representations based on deep learning. However, such representations are still vulnerable to emotion variability. To address this issue, we propose multiple improvements to train speaker encoders to increase emotion robustness. Firstly, we utilize CopyPaste-based data augmentation to gather additional parallel data, which includes different emotional expressions from the same speaker. Secondly, we apply cosine similarity loss to restrict parallel sample pairs and minimize intra-class variation of speaker representations to reduce their correlation with emotional information. Finally, we use emotion-aware masking (EM) based on the speech signal energy on the input parallel samples to further strengthen the speaker representation and make it emotion-invariant. We conduct a comprehensive ablation study to demonstrate the effectiveness of these various components. Experimental results show that our proposed method achieves a relative 19.29\% drop in EER compared to the baseline system.
\end{abstract}
\vspace{-0.1cm}
\vspace{-0.2cm}
\section{Introduction}
\label{sec:intro}

Speaker verification (SV) aims to determine if two speech samples come from the same person. Currently, SV systems that use low-dimensional speaker representations extracted from deep learning-based speaker encoders have become the dominant approach in this field. The performance of SV experienced a significant boost as researchers explored various neural network architectures \cite{snyder2018x, zeinali2019but, desplanques20_interspeech, zhang22h_interspeech}, experimented with diverse loss functions \cite{xiang2019margin, Chung2020, zhao2022multi, han2023exploring}, and implemented different back-ends \cite{ramoji2020nplda, wang22r_interspeech, zeng2022attention, brummer22_interspeech}. However, the development of SV systems capable of handling emotional speech is challenging due to the changes in acoustic features caused by emotional variations. These changes increase the intra-class variance of speaker representations, which can negatively impact the performance of SV systems.

Many studies have attempted to improve SV in emotional speech scenarios. It has been demonstrated that utterance-level speaker representations, such as i-vectors and x-vectors, are sensitive to emotions \cite{parthasarathy2017study, pappagari2020x}. To mitigate the variability of emotions within speakers, \cite{bao07_interspeech} introduced an emotion compensation strategy. In \cite{li20ba_interspeech}, emotion adversarial training was explored for suppressing the learning of information from various emotions. The authors in \cite{lertpetchpun23_interspeech} showed that replacing Batch Normalization (BN) layers with instance-based normalization layers in ResNet-like models reduces the models’ learning of emotional information. However, most previous studies have predominantly focused on small-scale datasets with a limited number of speakers, typically fewer than 100, which does not provide a solid foundation for reliable research. A lack of speaker diversity is widely known to result in poor model performance in real-world scenarios. Intuitively, a model can more effectively learn an emotion-invariant representation of a speaker when it has access to a large volume of parallel data, including recordings of the same speaker expressing various emotions. However, the collection of emotional parallel data is a time-consuming and costly endeavor. To our knowledge, few studies exist on the SV of emotional datasets with a large number of speakers due to data scarcity.

This paper focuses on the SV of emotional speech, involving a substantial number of speakers, and presents a scheme for learning emotion-invariant speaker representations. Dusha \cite{kondratenko23_interspeech} is a large open-source corpus for speech emotion recognition (SER) containing thousands of unique speakers. We perform some processing on the Dusha dataset to comply with our research requirements, and all experiments are conducted on this dataset to align with practical scenarios. Drawing inspiration from \cite{pappagari2021copypaste}, we utilize CopyPaste-based data augmentation to acquire more parallel data. Furthermore, we propose Emotion-invariant Representation Learning (ERL) that reduces the variation within speaker representations and minimizes their correlation with emotional information. ERL combines the AAM-Softmax \cite{deng2019arcface} loss and cosine similarity loss. The AAM-Softmax loss can boost SV performance by explicitly encouraging inter-class separability and intra-class compactness. The cosine similarity loss tries to keep the cosine similarity between a speaker’s two utterances with different emotion categories or intensities as large as possible. To further enhance ERL, we propose an emotion-aware masking (EM) strategy. Specifically, we use the energy of the speech signal to identify regions that are likely to contain emotional information and apply masking exclusively within these regions. The cosine similarity between speaker representations of emotional and unemotional utterances can be increased by combining EM and ERL. We carry out experiments to prove the effectiveness of the proposed method and conduct detailed ablation studies. 

\vspace{-0.2cm}
\section{LEARNING EMOTION-INVARIANT SPEAKER REPRESENTATIONS}
\label{sec:method}

\begin{figure}[t]
  \centering
  \includegraphics[width=\linewidth]{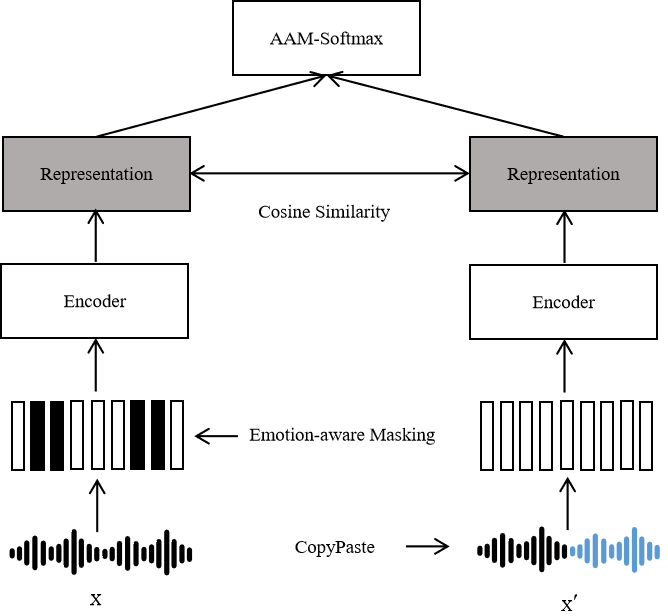}
    \caption{The proposed scheme for learning emotion-invariant speaker representations. Pairs of training samples are fed into a Siamese network with the shared parameter set. ${x}'$ is the result of applying CopyPaste to $x$.}
  \label{fig:schematic}
  \vspace{-0.5cm}
\end{figure}

In this section, we present a scheme for learning emotion-invariant speaker representations in detail. 
A diagram of the scheme can be found in Fig. \ref{fig:schematic}. For convenience, the data that fed into speaker encoder in the training stage are represented as $\left \{ \left ( {x}_{i},{y}_{i} \right ) \right \}_{i=1}^{N}$, where ${x}_{i}$ refers to input utterance i, ${y}_{i}$ refers to the speaker label of ${x}_{i}$, $N$ is the total number of training utterances. The utterance obtained by applying CopyPaste to ${x}$ is denoted as ${x}'$.

\vspace{-0.2cm}
\subsection{CopyPaste-based Data Augmentation}
\label{ssec:data augmentation}

CopyPaste is a simple yet effective data augmentation method originally developed for the SER task \cite{pappagari2021copypaste}. The core concept of this method is to create a new utterance by copying an existing one and pasting it either at the beginning or end of another utterance. 
To better suit the requirements of the SV task, we made several adjustments to the original CopyPaste technique. We argue that having a large amount of emotional parallel data can force the speaker encoders to focus more on the intrinsic characteristics of the speaker within the utterance under various emotional conditions. Our aim is to achieve more emotional diversity in parallel data without altering the speaker labels. The variability in speaker emotion not only manifests itself in different categories of emotion, but also in the presence of different levels of intensity for the same category of emotion. For example, the acoustic properties of happy and sad speech are significantly different, as well as hot and cold anger. We develop three schemes to accomplish this: 1) S-CP, where two utterances with the same emotion category and speaker label $y$ are concatenated to generate a new utterance with speaker label $y$ but possibly different emotion intensity; 2) D-CP, where two utterances with different emotion categories but the same speaker label $y$ are concatenated to generate a new utterance with speaker label $y$ and containing both emotion categories, and 3) S+D-CP, which combines both S-CP and D-CP.

\vspace{-0.2cm}
\subsection{Emotion-invariant Representation Learning}
\label{ssec:emotion-invariant learning}


We propose an objective function to aid AAM-Softmax in learning emotion-invariant speaker representations. Pairs of training samples, ${x}$ and ${x}'$, are processed by a Siamese network that shares a common set of parameters. To be specific, the process involves first extracting the acoustic features from the samples, which are then sent to the speaker encoder to obtain the speaker representations. After that, the AAM-Softmax and cosine similarity loss are calculated. The combined loss can be mathematically summarized as follows:
\begin{equation}
\footnotesize
\label{equ1}
  L_{ERL}= L_{AAM}\left ( x \right )+L_{AAM}\left ( {x}' \right )+\alpha L_{COS}\left ( x,{x}' \right )
\end{equation}
where $L_{AAM}$ is the AAM-Softmax loss, $L_{COS}$ is the cosine similarity loss, and $\alpha$ is a hyperparameter for weighted summation. The $L_{COS}$ is defined as:
\begin{equation}
\footnotesize
\label{equ2}
  L_{COS}\left ( x,x{}' \right )= -\sum_{i=1}^{N}\frac{v_{x}^{i}\cdot v_{x{}'}^{i} }{\left \| v_{x}^{i} \right \|\left \| v_{x{}'}^{i} \right \|}
\end{equation}
where $v_{x}^{i}$ and $v_{x{}'}^{i}$ represent the speaker representations for the i-th sample $x$ and ${x}'$, respectively. ${x}'$ is the result of applying CopyPaste to $x$. Even though both have the same speaker label, their emotions differ in terms of category or intensity. The $L_{COS}$ forces speaker encoder reduces the variance brought by emotions for each speaker and learns speaker representations that are invariant to emotions. SV systems typically use a simple cosine similarity scoring method to compare enrollment and test utterances. The use of cosine similarity loss during training aligns with the scoring method, thereby preventing a decrease in performance. 

\vspace{-0.2cm}
\subsection{Emotion-aware Masking}
\label{ssec:masking}

Empirical evidence suggests that the way ERL aligns utterances of varying emotional categories or intensities from the same speaker is suboptimal. It is hypothesized that aligning both emotional and unemotional utterances from the same speaker could lead to more emotion-agnostic speaker representations. To enhance ERL, an emotion-aware masking strategy is proposed. This strategy uses the root mean square (RMS) energy of the speech signal to determine mask locations. This energy, indicative of the speaker’s vocal loudness and intensity, can serve as a reflection of their emotional state \cite{el2011survey}. Typically, intense emotions such as anger or happy are associated with higher signal energy, while subdued emotions like sadness or depression correspond to lower signal energy.The RMS energy is defined as
\begin{equation}
\footnotesize
\label{equ3}
  RMS\left ( f \right )= \sqrt{\frac{1}{L}\sum_{l=1}^{L}\left | S_{f}\left ( l \right ) \right |^{2}}
\end{equation}
where $S$ is the input speech signal, $S_{f}$ denotes the f-th frame of $S$ with frame length $L$ and $RMS\left ( f \right )$ denotes the RMS energy of the f-th frame. 

The details of the EM process are as follows. First, we perform an utterance-level normalization by dividing the RMS energy by the maximum value, limiting it to $\left[0,1 \right ]$.
We then partition the energy into three zones: a high-energy zone with a range of $\left ( 0.5,1 \right ]$, a low-energy zone with a range of $\left ( 0.2,0.5 \right ]$, and a noise zone with a range of $\left [ 0,0.2 \right ]$. We count the number of frames in the high and low energy zones separately to determine which type of emotion dominates the voice. If there are more frames in the high energy zone than in the low energy zone, then the speech is dominated by intense emotions. Otherwise, it is dominated by subdued emotions. Subsequently, we randomly select $m$ mask positions from the dominant energy zone. These selected mask positions will become the centers of the masking regions, with each mask spanning $T$ frames. Once the emotional aspects of the speech are masked, it is considered to be devoid of emotion. Therefore, we make a simple improvement to the ERL training process, where the acoustic features of $x$ perform EM while ${x}'$ does not.

\vspace{-0.2cm}
\section{EXPERIMENTAL SETUP}
\label{sec:experiments}

\vspace{-0.2cm}
\subsection{Datasets}
\label{ssec:datasets}

Our models are first pre-trained on the VoxCeleb \cite{nagrani2020voxceleb} and then fine-tuned on the Dusha \cite{kondratenko23_interspeech} dataset to evaluate the performance of SV.

VoxCeleb is a dataset from YouTube with diverse speakers and speech in varied acoustic environments. It has two subsets: VoxCeleb1 and VoxCeleb2, mainly containing English speech data. The VoxCeleb2 development set is used for training, while the evaluation set is built from VoxCeleb1, including VoxCeleb1-O, VoxCeleb1-E, and VoxCeleb1-H.

Dusha is a large Russian speech dataset focused on SER that is composed of two subsets: Crowd and Podcast. The Crowd provides metadata with speaker identity and emotion category labels for each utterance, which is ideal for conducting our research. The original paper \cite{kondratenko23_interspeech} divides the Crowd into training and test sets, each containing four emotion categories: angry, positive, neutral, and sad. We use the Silero \footnote{https://github.com/snakers4/silero-vad} tool to remove silences from utterances in both sets. For the test set, we select utterances from the same speaker expressing four different emotions to create cross-emotion and same-emotion test trials. All test trials are merged to evaluate overall SV performance on emotional speech. The training and test sets follow the original paper’s division and with no speaker overlap. Table \ref{table1} lists the statistics of the training and test sets, while Table \ref{table2} lists the number of constructed test trails. The training sets, test sets, and test trails used in our experiments are available at https://github.com/campustian/LESRFSV.
\begin{table}[t]
\caption{The statistics on the of utterances, speakers, and duration}
\label{table1}
\centering
\setlength{\tabcolsep}{3mm}
\begin{tabular}{cccc}
\toprule
\textbf{Data}  & \textbf{Utterances} & \textbf{Duration}    & \textbf{Speakers} \\ 
\midrule
Train & 166,742    & 92.76 hours & 1,870    \\
Test  & 8,338      & 4.90 hours  & 84       \\ 
\bottomrule
\end{tabular}
\vspace{-0.5cm}
\end{table}
%
%
\begin{table}[t]
\caption{The number of test trails. A, P, N, and S denote angry, positive, neutral, and sad. Merged represents merging all the trails.}
\label{table2}
\centering
\setlength{\tabcolsep}{3mm}
\begin{tabular}{|c|cccc|}
\hline
       & \multicolumn{1}{c|}{A}      & \multicolumn{1}{c|}{P}      & \multicolumn{1}{c|}{N}      & S      \\ \hline
A      & \multicolumn{1}{c|}{44,496} & \multicolumn{1}{c|}{48,220} & \multicolumn{1}{c|}{45,534} & 44,580 \\ \hline
P      & \multicolumn{1}{c|}{48,220} & \multicolumn{1}{c|}{26,922} & \multicolumn{1}{c|}{47,182} & 42,034 \\ \hline
N      & \multicolumn{1}{c|}{45,534} & \multicolumn{1}{c|}{47,182} & \multicolumn{1}{c|}{43,874} & 45,902 \\ \hline
S      & \multicolumn{1}{c|}{44,580} & \multicolumn{1}{c|}{42,034} & \multicolumn{1}{c|}{45,902} & 32,378 \\ \hline
Merged & \multicolumn{4}{c|}{421,122}                                                                      \\ \hline
\end{tabular}
\vspace{-0.5cm}
\end{table}

\vspace{-0.2cm}
\subsection{Baseline System}
\label{ssec:baseline}

In our experiments, we utilize the ResNet34-TSDP \cite{wang2021revisiting} to serve as the speaker encoder. To increase the diversity of training samples, we apply three distinct augmentation techniques to the pre-trained model: 1) Noise augmentation, which includes noise, music, and babble, using the MUSAN \cite{snyder2015musan} corpus; 2) The addition of convolutional reverberation through the RIR \cite{ko2017study} Noise corpus; 3) Altering the tempo to either 0.9 times slower or 1.1 times faster, while maintaining a constant pitch. Our baseline model is fine-tuned using the AAM-Softmax, solely on the Crowd’s training set. We do not employ any data augmentation techniques in this process.


\vspace{-0.2cm}
\subsection{Training Details and Evaluation Metric}
\label{ssec:Training Details}

The acoustic features are 64-dimensional log Mel-filterbank energies, with a frame length of 25ms and a hop size of 10ms. These features are mean-normalized per utterance and randomly cropped to 2 seconds. All data augmentation is performed on-the-fly. And we use the CopyPaste-based method with segments that are 1 second long, resulting in lengths of 2 seconds.

All models are trained using the SGD optimizer with a weight decay of 2e-5. The batch size for training is 90, unless otherwise specified. During the pre-trained phase, the speaker encoder is trained using AAM-Softmax with a scale of 32. The margin is linearly increased during the first two epochs and then fixed at 0.25. The initial learning rate is 0.1 and is reduced by multiplying it by 0.9 after every 50,000 batches until the model converges. During the fine-tuned phase, models are trained for 100 epochs with a learning rate of 0.001. AAM-Softmax is used with a margin of 0.15 and a scale of 32. The hyper-parameters in the ERL and EM are set as follows: $\alpha=1$, $m=2$, and $T=7$. 

The cosine distance measurement is utilized for scoring trials. The Equal Error Rate (EER) is used to measure the performance of SV.

\vspace{-0.2cm}
\section{RESULTS}
\label{sec:results}
\begin{table}[t]
\caption{The EER of the pre-trained model on the VoxCeleb1.}
\label{table3}
\centering
\setlength{\tabcolsep}{0.6mm}
\begin{tabular}{cccc}
\toprule
\textbf{Model} & \textbf{Voxceleb1-O} & \textbf{Voxceleb1-E} & \textbf{Voxceleb1-H} \\
\midrule
ECAPA \cite{desplanques20_interspeech}      & \textbf{1.01\%}      & 1.24\%               & 2.32\%   \\
ResNet34-TSDP              & 1.07\%               & \textbf{1.21\%}      & \textbf{2.17\%}    \\
\bottomrule
\end{tabular}
\vspace{-0.5cm}
\end{table}
\begin{table*}[t]
\label{table4}
\caption{The performance of different models in terms of EER on the Crowd test set. A, P, N, and S denote angry, positive, neutral, and sad. Merged represents merging all the trails. Row 2 represents the baseline system. The Fine-tuned$^{\star}$ stage has a batch size that is double that of the Fine-tuned stage, due to the implementation of ERL training. RM stands for random masking.}
\resizebox{\linewidth}{!}{
\begin{tabular}{clccccccccccc}
\toprule
\multirow{2}{*}{\#} & \multirow{2}{*}{\textbf{Models}} & \multicolumn{4}{c}{\textbf{Same-emotion}} & \multicolumn{6}{c}{\textbf{Cross-emotion}} & \multirow{2}{*}{\textbf{Merged}} \\
\cmidrule(r){3-6} \cmidrule(r){7-12}
                    &                         & \textbf{A-A}    & \textbf{P-P}    & \textbf{N-N}    & \textbf{S-S}    & \textbf{A-P}     & \textbf{A-N}     & \textbf{A-S}     & \textbf{P-N}   & \textbf{P-S}   & \textbf{N-S}    &                         \\
\midrule
1                   & Pre-trained             & 8.31\% & 8.07\% & 9.12\% & 8.07\% & 12.05\% & 12.55\% & 14.92\% & 10.49\% & 11.96\% & 9.87\% & 11.23\%                 \\
2                   & Fine-tuned              & 4.45\% & 4.21\% & 4.14\% & 4.05\% & 5.73\%  & 6.40\%  & 8.01\%  & 5.53\%  & 6.41\%  & 4.95\% & 5.65\%                  \\
\midrule
                    & Fine-tuned               & \multicolumn{11}{c}{Comparative Study of Different Data Augmentation Techniques}                                  \\
                    \cmidrule(r){3-13}
3                   & + Noise                  & 4.38\% & 4.38\% & 4.03\% & 4.11\% & 5.51\%  & 6.47\%  & 7.67\%  & 5.77\%  & 6.50\%  & 4.70\% & 5.63\%                  \\
4                   & + Music                  & 4.30\% & 4.17\% & 4.15\% & 4.08\% & 5.61\%  & 6.38\%  & 7.88\%  & 5.53\%  & 6.70\%  & 4.90\% & 5.65\%                  \\
5                   & + Babble                 & 4.40\% & 4.22\% & 4.04\% & 4.33\% & 5.51\%  & 6.24\%  & 7.71\%  & 5.61\%  & 6.45\%  & 4.83\% & 5.60\%                  \\
6                   & + Tempo                  & 4.35\% & 4.32\% & 4.10\% & 4.15\% & 5.74\%  & 6.32\%  & 7.52\%  & 5.72\%  & 6.51\%  & 4.88\% & 5.59\%                  \\
7                   & + S-CP                   & 4.14\% & 4.29\% & 3.86\% & 3.81\% & 5.63\%  & 6.07\%  & 6.99\%  & 5.62\%  & 5.95\%  & 4.45\% & 5.27\%                  \\
8                   & + D-CP                   & 3.94\% & 4.27\% & 3.83\% & 3.55\% & 5.52\%  & 6.09\%  & 7.08\%  & 5.36\%  & 5.99\%  & 4.51\% & 5.28\%                  \\
9                   & + S+D-CP                 & 3.93\% & 4.11\% & 3.98\% & 3.79\% & 5.16\%  & 6.05\%  & 7.06\%  & 5.50\%  & 6.00\%  & 4.57\% & 5.27\%                  \\
\midrule
                    & Fine-tuned$^{\star}$     & \multicolumn{11}{c}{S-CP Based Ablation Study}                                  \\
                    \cmidrule(r){3-13}
10                  & + S-CP                   & 3.89\% & 4.09\% & 4.02\% & 3.89\% & 5.39\%  & 6.07\%  & 7.05\%  & 5.46\%  & 5.87\%  & 4.69\% & 5.25\%                  \\
11                  & ++ ERL                   & 3.64\% & 3.68\% & 3.87\% & 3.95\% & 4.82\%  & 5.49\%  & 6.31\%  & 4.93\%  & 5.58\%  & 4.17\% & 4.87\%                  \\
12                  & +++ RM                   & 3.85\% & 3.64\% & 3.53\% & \textbf{3.37\%} & 5.13\%  & 5.64\%  & 6.46\%  & 4.90\%  & 5.13\%  & 4.00\% & 4.81\%         \\
13                  & +++ EM                   & \textbf{3.52\%} & 3.78\% & 3.82\% & 3.61\% & 4.89\%  & 5.40\%  & 6.23\%  & 4.99\%  & 5.15\%  & 3.98\% & 4.80\%         \\
\midrule
                    & Fine-tuned$^{\star}$      & \multicolumn{11}{c}{S+D-CP Based Ablation Study}                                  \\
                    \cmidrule(r){3-13}
14                  & + S+D-CP                 & 4.02\% & 4.12\% & 3.89\% & 4.10\% & 5.57\%  & 6.13\%  & 6.85\%  & 5.53\%  & 5.67\%  & 4.60\% & 5.24\%                  \\
15                  & ++ ERL                   & 3.60\% & 3.82\% & 3.70\% & 3.49\% & 4.90\%  & 5.51\%  & 6.12\%  & 4.87\%  & 5.05\%  & 3.95\% & 4.70\%                  \\
16                  & +++ RM                   & 3.57\% & \textbf{3.61\%} & 3.96\% & 3.68\% & \textbf{4.77\%}  & 5.56\%  & 5.93\%  & 4.97\%  & 5.13\%  & 4.08\% & 4.75\%                  \\
17                  & +++ EM                   & 3.64\% & 3.95\% & \textbf{3.46\%} & 3.40\% & 4.90\%  & \textbf{5.23\%}  & \textbf{5.86\%}  & \textbf{4.76\%}  & \textbf{4.94\%}  & \textbf{3.83\%} & \textbf{4.56\%}                  \\
\bottomrule
\end{tabular}
}
\vspace{-0.5cm}
\end{table*}

\vspace{-0.2cm}
\subsection{Performance of the baseline system}
\label{ssec:baseline results}

The performance of the pre-trained model on VoxCeleb1 is presented in Table \ref{table3}, with the results of the ECAPA model referenced from \cite{desplanques20_interspeech}. On the VoxCeleb1-O and VoxCeleb1-E test sets, the performance of ResNet34-TSDP is comparable to that of ECAPA. Additionally, a relative improvement of 6.5\% is observed on the VoxCeleb1-H test set. The performance of the pre-trained model on the Crowd test sets is shown in row 1 of Table \ref{table4}, where a sharp drop in performance is evident. A potential explanation for this difference in performance could be the language mismatch between VoxCeleb and Crowd. 
As indicated in row 2 of Table 4, which corresponds to our baseline system, the model’s EER saw a substantial improvement after fine-tuning. Specifically, it dropped from 11.23\% to 5.65\% on the Merged test set. However, a noticeable performance gap between the VoxCeleb1 and Crowd test sets still exists. We believe this discrepancy is due to the average duration of utterances in the Crowd test set being approximately 2 seconds, in contrast to the VoxCeleb1 set, which averages around 8 seconds. As expected, emotional speech degraded SV performance. The baseline model’s performance on cross-emotion test trials is generally 0.9\% to 3.96\% worse in absolute terms than on same-emotion test trials.

\vspace{-0.2cm}
\subsection{Performance of the proposed system}
\label{ssec:proposed method results}

Table \ref{table4} outlines the results of our proposed systems in rows 7 to 17. Specifically, rows 7 to 9 detail the model’s outcomes when trained with CopyPaste-based data augmentation and a batch size of 90, while rows 10 to 17 show the results with the same augmentation but a batch size of 180, as required by ERL for paired samples. The performances of rows 7 and 9 are comparable to those of rows 10 and 14, suggesting that the batch size does not significantly influence the results. Due to the scarcity of cross-emotional parallel data for some speakers in the Crowd training set, ERL-related experiments were confined to S-CP and S+D-CP. A comparison of row 13 with row 2 shows that integrating S-CP with ERL and EM reduces the EER from 5.65\% to 4.80\% on the Merged test set. Similarly, comparing row 17 with row 2 reveals that combining S+D-CP with ERL and EM lowers the EER from 5.65\% to 4.56\%. As indicated in rows 14, 15, 17, and 2, S+D-CP contributes to a relative improvement of 7.26\%. The additions of ERL and EM further enhance these improvements by 9.55\% and 2.48\%, respectively, culminating in a total relative improvement of 19.29\%. Furthermore, the performance gap between cross-emotion and same-emotion test trials has been reduced. Notably, the EER performance of the N-S is now lower than that of the P-P test set.

\vspace{-0.2cm}
\subsection{Ablation studies}
\label{ssec:comparative results}

\vspace{-0.2cm}
\subsubsection{Comparison of different data augmentation methods}
\label{ssec:comparision with other data augmentation methods}

This section compares four prevalent data augmentation techniques (adding noise, music, babble, and tempo change) with the proposed CopyPaste-based method. The augmentation is applied on-the-fly with a 50\% probability. Rows 3 to 9 of Table \ref{table4} show the results. Compared to the baseline in row 2, the four commonly used data augmentation techniques do not improve the SV in emotional speech. However, the three CopyPaste-based data augmentation techniques demonstrate comparable performance, with the EER decreasing from 5.65\% to approximately 5.27\% on the Merged test set, resulting in a relative improvement of 6.73\%. These results suggest that constructing parallel data allows the speaker encoder to learn representations with reduced emotional content.

\vspace{-0.2cm}
\subsubsection{Effectiveness of the ERL}
\label{ssec:effectiveness of ERL}

We compare the performance of ERL combined with S-CP and ERL combined with S+D-CP, their results are shown in rows 10 to 11 and rows 14 to 15 of Table \ref{table4}, respectively. Paired training samples were constructed using S-CP, and ERL was then applied to decrease the EER from 5.25\% to 4.87\% on the Merged test set, achieving a relative improvement of 7.23\%. These results show that S-CP constructs speech with the same emotion category but different emotion intensity. At the same time, the ERL makes the speaker representation insensitive to speaker emotion intensity, thereby improving SV performance. When S+D-CP was used to construct paired training samples, ERL could further reduce the EER from 5.24\% to 4.70\%, resulting in a relative improvement of 10.31\%. The results demonstrate the effectiveness of ERL, with improved performance achieved through the construction of more diverse parallel data.

\vspace{-0.2cm}
\subsubsection{Effectiveness of the EM}
\label{ssec:effectiveness of EM}

In contrast to the random masking (RM) strategy, the proposed EM can enhance ERL performance. Rows 12 to 13 and rows 16 to 17 of Table \ref{table4} compare these two masking strategies. To ensure a fair comparison, the settings for RM were kept consistent with those used in EM. On the Merged test set, when S-CP was used to construct paired training samples, RM and EM achieved comparable performance, with only a slight improvement over no masking strategy (row 11). However, when S+D-CP was used to construct paired training samples, RM slightly deteriorated the performance, and the EER increased from 4.70\% to 4.75\%. Conversely, EM showed a clear advantage, and the EER decreased from 4.70\% to 4.56\%, resulting in a 2.98\% relative improvement compared to no masking strategy (row 15).

\vspace{-0.2cm}
\section{CONCLUSIONS}
\label{sec:conclusion}

In this paper, we propose a scheme for learning speaker representations that are invariant to emotions. We first verified that emotional utterances degrade SV performance, with cross-emotion test trails performed worse than same-emotion test trails. Then to alleviate this problem, we propose three improved techniques to train the speaker encoder. By using CopyPaste-based data augmentation, more parallel data for emotions can be created, resulting in a 7.26\% relative performance improvement. Additionally, ERL can be used to minimize the variance within a speaker caused by emotions, leading to an additional 9.55\% relative performance improvement. EM can further enhance ERL performance by masking emotional regions based on the energy of speech, resulting in a 2.48\% relative performance improvement. Overall, the proposed method achieves a 19.29\% relative EER reduction compared to the baseline system.

\vfill\pagebreak
\ninept
\bibliographystyle{IEEEbib}
\bibliography{strings,refs}

\end{document}